# Jordan Journal of Physics

## ARTICLE

## Magnetization Measurements on "as Prepared" and Annealed $Fe_{3-x}Mn_xSi$ Alloys

**M. S. Lataifeh**[a], **M. O'Shea**[b], **A. S. Saleh**[c] and **S. H. Mahmood**[c]

[a] *Department of Physics, Mu'tah University, Mu'tah, Jordan*
[b] *Kansas State University, Manhattan, KS 66506-2601, USA*
[c] *Department of Physics, Yarmouk University, Irbid, Jordan*

**Abstract:** The magnetic properties of the alloy system $Fe_{3-x}Mn_xSi$ have been studied by measuring magnetization for samples with x = 0, 0.1, 0.25, 0.5, and by thermal scanning techniques for samples with x = 0, 0.1. The results reveal that the system is ferromagnetic in this composition range. Zero field cooling and field cooling magnetization measurements indicate a similar magnetic ordering and magnetic anisotropy in all samples. The saturation magnetization for the annealed samples was higher than that for "as prepared" samples. This is attributed to the reduction of magnetic domain boundaries rather than to improving magnetic order as a result of annealing. Further, $T_C$ values determined from thermal DSC measurements are in good agreement with previously reported results based on magnetic measurements.
**Keywords:** Heusler Alloys, Magnetization, Magnetic Order.

## Introduction

The unit cell of the $Fe_{3-x}Mn_xSi$ alloys consists of four interpenetrating f.c.c sublattices A, B, C, and D, with origins displaced along the body diagonal at the points A (0,0,0), B (1/4,1/4,1/4), C (1/2,1/2,1/2) and D (3/4, 3/4, 3/4). $Fe_3Si$ crystallizes in the $DO_3$ structure with the A, B, and C sites occupied by Fe and the D site by Si [1, 2].

The magnetic behavior of the alloy $Fe_{3-x}Mn_xSi$ exhibits ferromagnetism at low Mn concentration (x<0.75), and shows a complex magnetic structure at higher values of x [1]. As Mn is introduced into the alloy, it preferentially occupies the B site for x< 0.75, and then begins to occupy the A, C sites as x increases. It has been found also that nearest neighbor interactions dominate at low Mn concentration [3].

This system has attracted the interest of many researchers [1-12] since it shows a variety of magnetic phases as x is increased up to 3.0. It is ferromagnetic below x=0.75, and shows spin glass behavior for 0.75< x <1.2.

The substitution of Mn for Fe in these pseudo-binary alloys has been studied using various techniques. Fe Mössbauer spectroscopy (MS) was used to study the dependence of the average hyperfine fields at the different sites on Mn concentration [2-4, 7, 8, 12]. Magnetic and crystallograghic properties have also been studied by X-ray and neutron diffraction techniques, together with magnetization, resistivity, and magnetostriction measurements [1,9]. Ferromagnetic resonance studies have also been made on this system [5,6]. The above results show that the Mn substitution reduces the magnetic exchange energy, thus leading to a reduction in the critical (Curie or Neel) temperature.

In this work we present the results of magnetization measurements using the Superconducting Quantum Interference Device (SQUID), and the dependence of magnetization on the temperature under zero field cooling and field cooling conditions

**Corresponding Author:** Mahdi S. Lataifeh. **Email:** mahdi_lataifeh@yahoo.com

on the as prepared and annealed alloys. Also, thermal measurements using Differential Scanning Calorimetry (DSC) is used to determine the Curie temperature for some of the samples under investigation. This work is part of an ambitious project, intended to carry a thorough investigation of the system by studying a series of samples with different Mn concentrations using MS and X-ray techniques.

**Experimental**

Appropriate proportions of high purity metals (better than 99.9%) were melted in an argon arc furnace, forming a series of the alloy buttons. The buttons were flipped and remelted several times in order to insure homogeneity. The mass of each button was measured and compared with the total mass of the initial powders; mass losses of 1% or less were found, which indicates that the achieved concentrations are consistent with the desired ones. The alloy buttons were then crushed into fine powders, part of which was used to prepare the "as prepared" samples. The "annealed samples" were prepared by annealing parts of the powders under vacuum in quartz tubes for two weeks at 800 °C and then left to cool to room temperature.

Magnetization measurements were performed using a superconducting Quantum Interference Device (SQUID) Magnetometer, at the Department of Physics, Kansas State University, USA. Measurements were made in the temperature range from 10 K to 400 K in an applied magnetic field up to 5.5 Tesla. To reduce demagnetization, the powders were placed in cylindrical sample holders.

For thermal DSC measurements, 25-40 mg of the powder were placed in a crucible beside an inert reference and subjected to the same temperature program. The temperature was scanned from 100 °C to 600 °C. The magnetic phase transition could be exothermic or endothermic, which produces a signal in the DSC scan, from which the critical transition temperature is determined.

**Results and Discussion**

Fig. 1 shows the Zero Field Cooled (ZFC) and the Field Cooled (FC) magnetization curves for the two samples $Fe_3Si$ and $Fe_{2.75}Mn_{0.25}Si$. The ZFC curves were obtained by cooling the sample from Room Temperature (RT) down to 10 K in zero field, and then a small field (200 Oe) is applied and the magnetization is measured as the sample is warmed up to 400 K. This procedure will have a maximum effect on the magnetic domains with the least magnetic anisotropy, and will produce a net magnetization in the direction of the applied field.

The FC curves were measured by first cooling the sample down to 10 K in the presence of the small applied field (200 Oe) and then measuring the magnetization as the sample is warmed up to 400 K. Cooling down the sample in a 200 Oe field is sufficient to have more alignment of magnetic domains at 10 K. So we expect to observe larger magnetization in the FC process than in ZFC case.

The curves (Fig. 1) show almost constant magnetization as the temperature increases, indicating ferromagnetic order. The difference in magnetization between the ZFC and the FC processes is small and similar for the samples under investigation, indicating similar magnetic anisotropy in this composition range. The observed ferromagnetic behavior is consistent with the results of other workers using magnetization [1,9] and MS [4,7,8,12] measurements. Fig.1 also reveals a reduction in the magnetization upon replacement of Fe by Mn, which indicates a lower magnetic moment per formula unit. These findings are consistent with the observed decrease in the average hyperfine field with increasing Mn concentration as obtained from Mössbauer spectra [8,12].

Magnetization measurements on "as prepared" and annealed $Fe_{3-x}Mn_xSi$ alloys

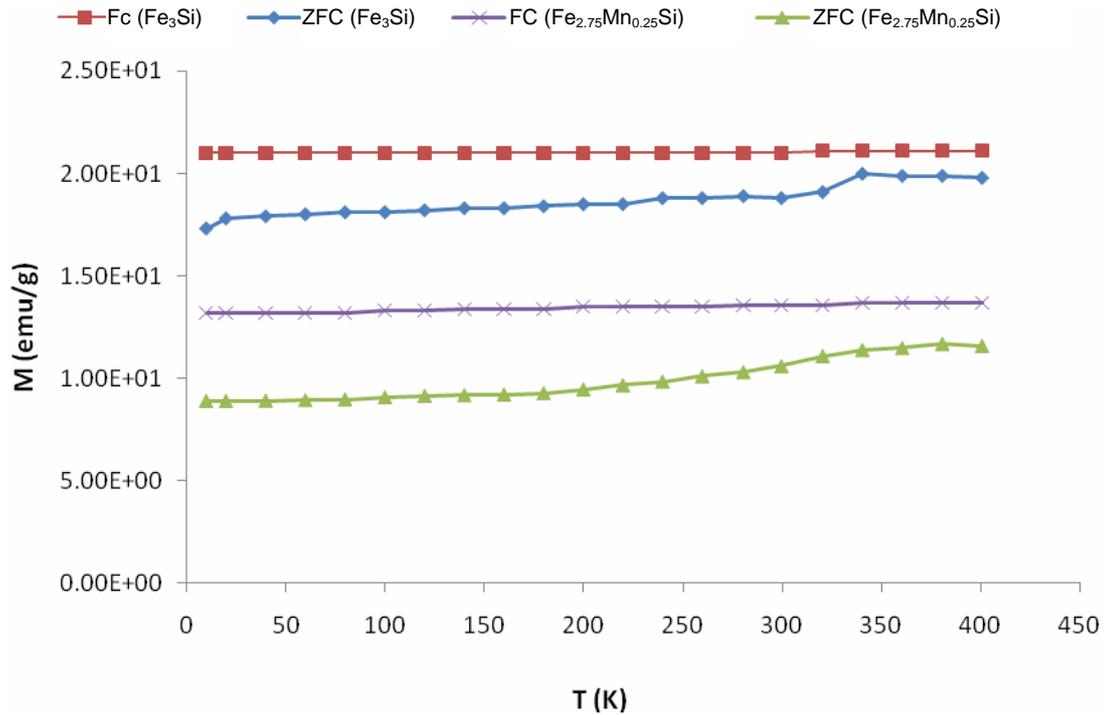

**Fig. 1:** ZFC and FC magnetization curves for the samples $Fe_3Si$ and $Fe_{2.75}Mn_{0.25}Si$. The solid line is a guide for the eye.

The spontaneous magnetization at 10 K was found to be higher for FC process than for ZFC. This indicates that the partial alignment of the moments due to the application of the 200 Oe field in the FC process is responsible for the difference in magnetization M [M(FC) – M(ZFC) > 0]. This difference becomes smaller as T increases due to the slight increase in M in the ZFC process, and the saturated value of M in the FC process. The increase in M in the ZFC process is associated with progressively increasing the freedom for the orientation of the magnetic moments in the direction of the applied field due to the progressive unblocking of the frozen random state at 10 K. Ideally, the ZFC and FC curves would meet at 400 K. However, due to the irreversibility of the magnetic properties upon cooling and heating, a small difference in M is observed at 400 K. It is obvious also that M does not decrease with increasing temperature, indicating that the temperature range of the measurements is way below $T_C$.

The saturation magnetization, $M'_s$ for the annealed samples is higher than that for the as prepared ones at a given temperature (Fig.2). This result could be attributed to either the improvement of magnetic order, or the reduction in the magnetic domain boundaries as a result of annealing. However, MS data [8] do not show any appreciable difference in the hyperfine field distributions for the two sets of samples. Due to the fact that MS is sensitive to local atomic environments, but not to domain structure, it is more likely that the effect of annealing results in the reduction of domain boundaries rather than decreasing the magnetic disorder.

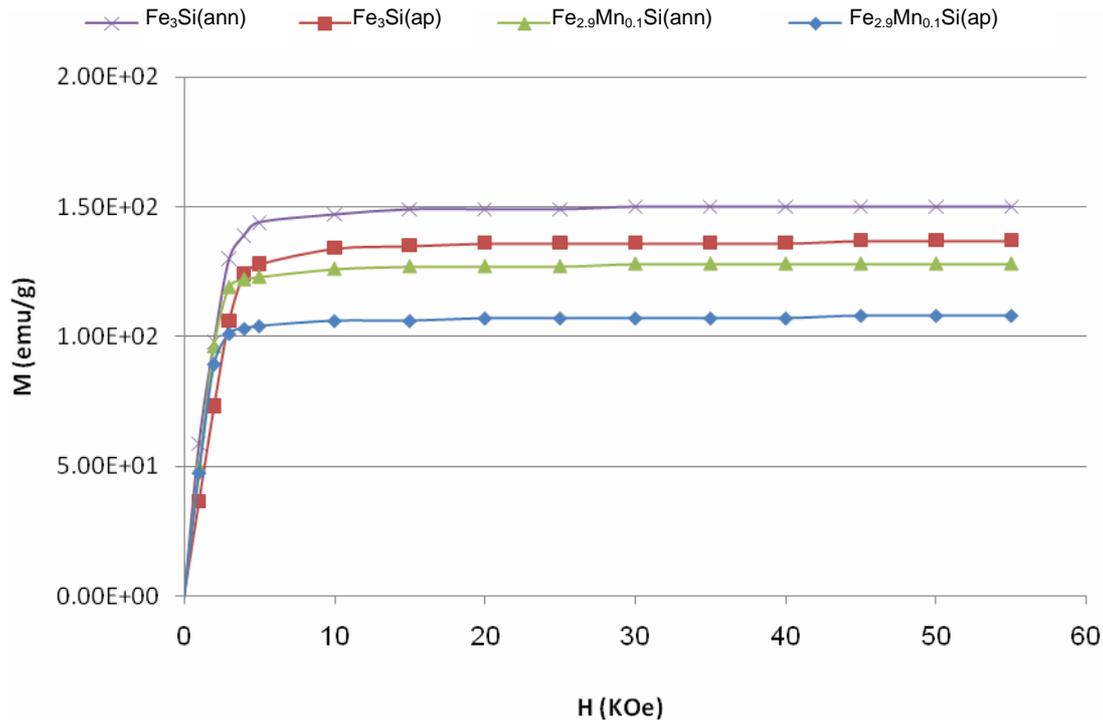

**Fig. 2:** Magnetic isothermal curves (M vs H) at 300 K for annealed $Fe_3Si$, as prepared $Fe_3Si$, annealed $Fe_{2.9}Mn_{0.1}Si$, and as prepared $Fe_{2.9}Mn_{0.1}Si$.

Also, Fig.3 shows the magnetization curves for the annealed samples at 300 K and at 10 K. The figure shows that the difference in $M_s$ between RT and 10 K for a given sample is small (~6%). This small change in $M_s$ is due to the fact that the measurements were made far below the Curie temperature, $T_C$ (RT $\leq 0.5\ T_c$). Fig.3 also shows a reduction in the saturation magnetization with increasing x, consistent with the reduction of the magnetic moment per formula unit upon replacement of Fe by Mn.

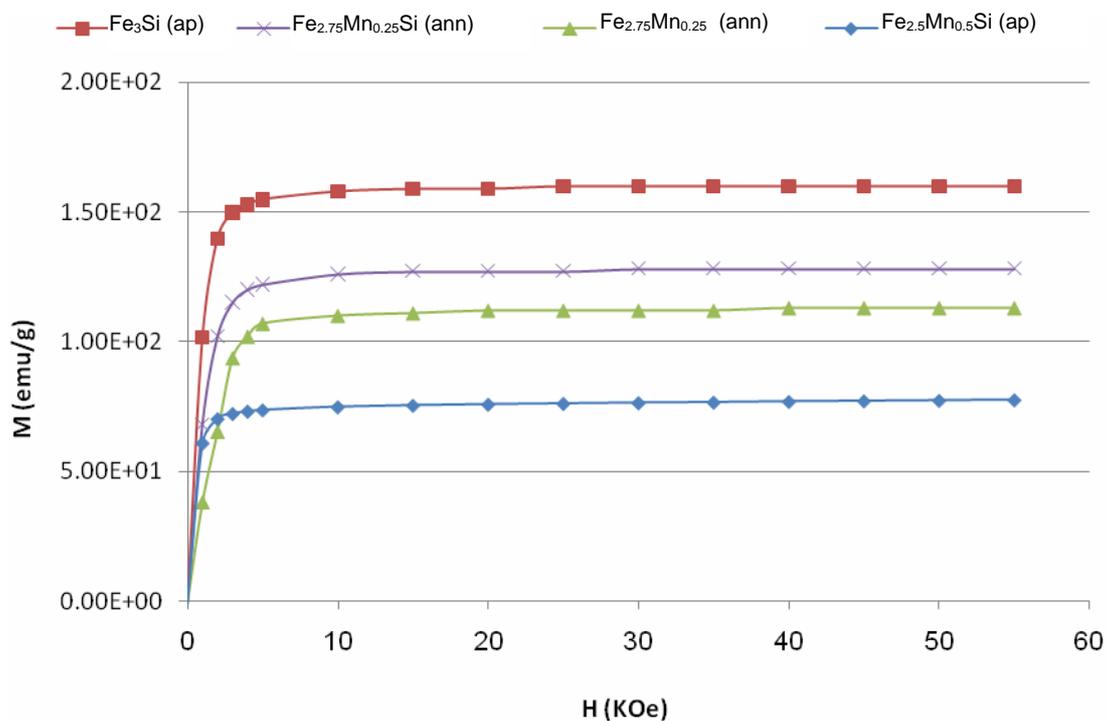

**Fig. 3:** Magnetic isothermal curves (M vs H) for as prepared $Fe_3Si$ at 10 K, annealed $Fe_{2.75}Mn_{0.25}Si$ at 10 K, annealed $Fe_{2.75}Mn_{0.25}Si$ at 300 K, as prepered $Fe_{2.5}Mn_{0.5}Si$ at 300 K.

Differential scanning calorimetry (DSC) measurements were performed for the two samples Fe$_3$Si and Fe$_{2.9}$Mn$_{0.1}$Si. The Curie temperature for these samples deduced from the measurements were 849 K and 794 K, respectively. It was reported in reference [1] that T$_C$ for the system under investigation decreases almost linearly as x increases up to 0.9. Fig.4 shows that our values for T$_C$ determined from the DSC measurements are consistent with the linear drop of T$_C$ with increasing x up to 0.9. Our results are also in good agreement with previous results [1,3,10, 11].

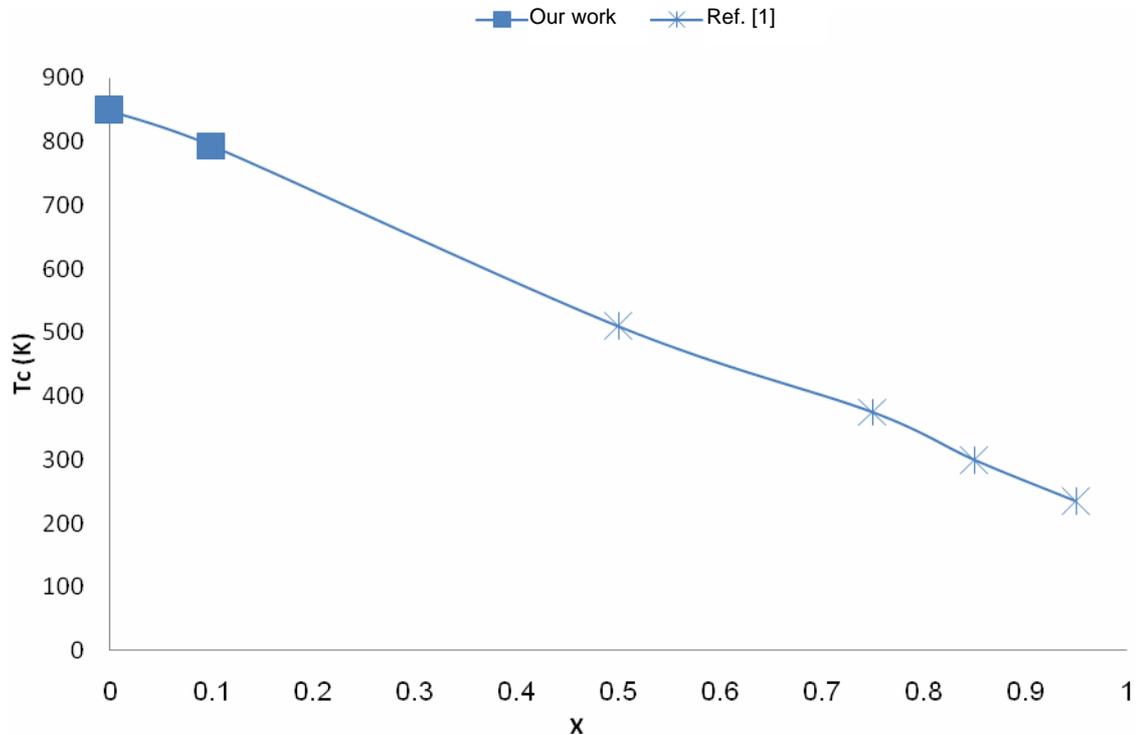

**Fig. 4:** Variation of the Curie temperature T$_C$ with composition x (this work for x= 0.0 and 0.1, and other values are from reference [1]). The solid line is a guide for the eye.

**Conclusions**

Thermal magnetization curves for samples of Fe$_3$Si and Fe$_{2.75}$Mn$_{0.25}$Si have shown difference in M between the ZFC and the FC processes. The magnetization measured in the ZFC process increase slightly with increasing the temperature due to the progressive unblocking of the magnetic moments, while that measured in the FC process remains almost constant with increasing temperature.

The saturation magnetization at a given temperature is found higher for the annealed sample than for the as prepared one. This is attributed to the reduction of the magnetic domain boundaries rather than the improvement of magnetic order as a result of annealing.

The Curie temperature T$_C$ was determined for the samples with x = 0 and 0.1 using DSC measurements. Our results are consistent with previous results obtained from magnetization measurements.

**Acknowledgment**

One of the authors (M.S. Lataifeh) would like to acknowledge the grant obtained from the Center for Theoretical and Applied Physical Sciences (CTAPS) at Yarmouk University as a partial support towards the completion of this work. Thanks are due to Dr. Ali Qudah from Mu'tah university for his help in the DSC measurements.